\documentclass[12pt]{article}

\usepackage{graphicx}

\newsavebox{\sboxpubnumber}

\newsavebox{\sboxpubdate}

\newcommand{\pubdate}[1]{\begin{lrbox}{\sboxpubdate}{#1}\end{lrbox}}

\newcommand{\Title}[1]{\begin{center} {\Large #1 } \end{center}}

\newcommand{\Author}[1]{\begin{center}{ \sc #1} \end{center}}

\newcommand{\Address}[1]{\begin{center}{ \it #1} \end{center}}

\newenvironment{Abstract}{\begin{quotation}  }{\end{quotation}}

\newenvironment{Presented}{\begin{quotation} \begin{center}

             PRESENTED AT\end{center}\bigskip

      \begin{center}\begin{large}}{\end{large}\end{center}

      \end{quotation}}

\newcommand{\Acknowledgements}{\bigskip  \bigskip \begin{center} \begin{large}

             \bf ACKNOWLEDGEMENTS \end{large}\end{center}}

\def\rmd{{\rm d}}
\def\F{{\rm F}}
\def\SUSY{{\rm SUSY}}
\def\halo{{\rm halo}}
\def\ev{\,{\rm eV}}
\def\baryon{{\rm baryon}}

\begin{document}






\begin{titlepage}

\pubdate{\today}                    


\vfill

\Title{The Ultra High Energy Cosmic Ray Spectrum from Relic Particle
  Decay and DGLAP Evolution}

\vfill

\Author{Ramon Toldr\`a}

\Address{Theoretical Physics, University of Oxford \\

         1 Keble Road, Oxford OX1 3NP, UK}

\vfill







\begin{Abstract}

We calculate the spectrum of ultra high energy cosmic rays produced by
the decay of a superheavy dark matter population clustered in the
galactic halo. To perform this calculation we start with fragmentation
functions measured at LEP and evolve them to the cosmic ray energy
scale using the QCD DGLAP equations. We consider Standard Model
evolution and supersymmetric evolution. We also take into account
many--body final states in the decay of the dark matter particles.
\end{Abstract}

\vfill

\begin{Presented}

    COSMO-01 \\

    Rovaniemi, Finland, \\

    August 29 -- September 4, 2001

\end{Presented}

\vfill

\end{titlepage}

\def\thefootnote{\fnsymbol{footnote}}

\setcounter{footnote}{0}




\section{Motivation}

Ultra High Energy Cosmic Rays (UHECR) are microscopic particles
--protons, photons or perhaps more exotic objects-- with a macroscopic
energy, about 50 Joules per particle in the extreme of the
present observed spectrum. They strike the upper layers of the Earth
atmosphere at a rate of about one event per century and per kilometer
square. Over one hundred of them have been detected so far by
kilometer scale detectors and many more are expected to be seen by
forthcoming observatories.

Explaining how these particles get this huge energy is a real
challenge for our present understanding of the cosmos. One UHECR
primary gets somehow ten million times the energy that a proton will
gain at the future Large Hadron Collider (LHC) at CERN. Nature has
always been at ease beating the achievements of mankind. One can
envisage two broad classes of mechanisms by which nature can produce
UHECRs, the so-called bottom up class models and the top-down class
models.

In bottom-up models charged particles are accelerated by magnetic
fields in large astrophysical sites. In top-down models particles are
not accelerated but are created at birth with the huge energy typical
of UHECRs.

Here we will concentrate on a particular top-down 
model~\cite{chor,bkv97,bs98}. We will
assume the UHECRs are produced by the decay of a population of
superheavy dark matter particles with a lifetime longer than the age of
the Universe $\tau > 10^{10}$~yr and with mass $\sim 100$ Joule$/c^2$.
Theoretical motivation for these particles can be found in~\cite{crypton}.

A superheavy dark matter population created at some stage of the early
universe~\cite{gravprod,reheatprod} will gravitationally cluster in
the galactic halo. Since the length scale of the halo is around 100
kpc, UHECRs produced in the galactic halo will not have time to
interact with the CMB before they reach the Earth. The absence of GZK
cut-off is a genuine prediction of models where UHECRs are produced by
the decay of a superheavy dark matter particle clustered in the
galactic halo.

Since the halo shape is close to spherical one expects a quasi
isotropic distribution of events from a halo superheavy dark matter
population, which is compatible with experiments. At present it is not
possible to make any strong claim about the observed angular
distribution of UHECR events because of low statistics. Hopefully,
future observatories like Pierre Auger will gather a large enough
sample of events to settle down this issue~\cite{efs01}.

Finally there is the question of the UHECR composition. The cosmic ray
observatories cannot measure the composition of the primary flux
(whether they are photons, protons or heavy nuclei) in an event by
event basis. Composition can only be determined in a statistical way.
Present analyses tend to favour protons as the main component in the
primary flux~\cite{ahvwz00}. In top-down
models the main component in the production site, in our case the the
galactic halo, is the photon component (actually, neutrinos dominate
over photons in a large range of the spectrum but their probability to
be detected is too small with present or past detectors). However, on
their way to the Earth, photons will interact with the low frequency
radio background in the Galaxy and their total flux may be
substantially diminished so that on the Earth the total photon flux
may be comparable or smaller than the baryon flux. Whether this is
indeed possible, subject to the EGRET bound on the low energy
$\gamma$-rays which will be created by the electromagnetic cascading,
will be discussed elsewhere.

Summing up, there are three main tests to falsify or support the
hypothesis of UHECRs produced by the decay of dark matter particles
clustered in the galactic halo. The first one is the energy
distribution of events or spectrum. The second one is the expected
angular distribution of events in the sky. The third one is the
composition of the primary flux. Here we will focus on the first test
and, partially, on the third one. We will briefly show how quantum
chromodynamics (QCD) can be used to calculate the spectrum and
composition (without photon galactic processing) of the expected UHECR
flux. For further details see~\cite{SarkarToldra,Toldra}. QCD was also
used to calculate the spectra of UHECRs
in~\cite{Rubin,FodorKatz,BerezKachel01}.

\section{DGLAP Evolution} \label{FFEvolution}

Let us assume that a dark matter particle $X$ population, superheavy
($M_X> 3\times 10^{12}$~GeV) and metastable, with a lifetime larger
than the age of the universe, is clustered in the galactic halo.
Barring unnatural relations between the coupling constant of $X$ and
the other fields the total particle multiplicity produced by its decay
will be dominated by partonic decays. The flux for the primary $h$
(baryon, photon or neutrino) is proportional to the inclusive decay
width
\begin{equation} 
 \label{eq:DecayRate}
J^h(E) \propto 
 \frac{1}{\Gamma_X}\frac{\rmd \Gamma(X \rightarrow h + \dots)}{\rmd x} = 
  \sum_a \int^1_x \frac{\rmd z}{z}\, \frac{1}{\Gamma_a}
  \left.\frac{\rmd \Gamma_a(y,\mu^2,M_X^2)}{\rmd y}\right|_{y=x/z} 
  D^h_a (z,\mu^2).
\end{equation}
Particle $h$ carries a fraction $x$ of the maximum available momentum
$M_X/2$, and a fraction $z$ of the parton $a$ momentum. The first term
in the integrand is the decay width of $X$ into parton $a$,
$\rmd\Gamma_a/\rmd\,y$, which is calculable in perturbation theory; in
lowest order and for 2-body decay it is proportional to $\delta(1-y)$.
The second factor, the non-perturbative $D^h_a$, is the fragmentation
function (FF) for particles of type $h$ from partons of type $a$. It
gives the expected mean number of particles $h$ coming from parton $a$.
Fragmentation functions cannot be calculated from firsts principles
but their dependence in the energy scale $\mu$ is governed by the
Dokshitzer--\-Gribov--\-Lipatov--\-Altarelli--\-Parisi (DGLAP)
equations~\cite{AltarelliParisi,DGL}.
\begin{equation}
 \label{eq:AP}
 \frac{\partial D^h_a(x,\mu^2)}{\partial\ln\mu^2} = 
 \sum_b \frac{\alpha_s(\mu^2)}{2\pi} 
 P_{ba}(x,\alpha_s(\mu^2)) \otimes D^h_b(x,\mu^2),
\end{equation}
where $\alpha_s(\mu^2)$ is the strong coupling constant and
$P_{ba}(x,\alpha_s)$ is the splitting function for the parton
branching $a\rightarrow\,b$. Here the convolution of two functions
$A(x)$ and $B(x)$ is defined as
\begin{equation}
 \label{eq:Convolution}
 A(x) \otimes B(x) \equiv \int^1_x \frac{\rmd z}{z}\, A(z) B(\frac{x}{z}).
\end{equation}
The splitting functions can be expanded perturbatively:
\begin{equation}
 \label{eq:LO}
 P_{ba}(x,\alpha_s) = P_{ba}(x) + {\cal O}(\alpha_s).
\end{equation}
We limit our study to leading order in $\alpha_s$ and therefore ignore
${\cal O}(\alpha_s)$ corrections to the splitting functions. It is
also convenient to define the following dimensionless evolution
parameter
\begin{equation}
 \label{eq:Tau}
 \tau \equiv \frac{1}{2\pi b}\ln\frac{\alpha_s(\mu^2_0)}{\alpha_s(\mu^2)}\ ,
\end{equation}
$b$ being the coefficient in the leading order $\beta$-function
governing the running of the strong coupling:
$\beta(\alpha_s)=-b\alpha_s^2$. We take $D^h_a$ to represent the
sum of particle $h$ and, if different, its antiparticle $\bar{h}$

In the Standard Model there are two partons species: gluons $g$ and
quarks $q$ (for the sake of brevity we shall only consider the singlet
quark, i.e., the sum of all quark and antiquark flavours). The DGLAP
equations can be written as 
\begin{equation}
  \label{eq:SM2X2}
  \partial_\tau 
  \left(
    \begin{array}{l}
      D_q \\
      D_g
    \end{array}
  \right)
  =
  \left(
    \begin{array}{cc}
      P_{qq} & 2n_{\F}P_{gq} \\
      P_{qg} & P_{gg}
    \end{array}
  \right)
  \otimes
  \left(
    \begin{array}{l}
      D_q \\
      D_g
    \end{array}
  \right).
\end{equation}

When supersymmetry (SUSY) is included one has in addition the gluinos
and the squark singlet. The DGLAP equations are now
\begin{equation}
  \label{eq:SUSY4X4}
 \partial_\tau 
  \left(
    \begin{array}{l}
      D_q \\
      D_g \\
      D_s \\
      D_\lambda
    \end{array}
  \right)
  =
  \left(
    \begin{array}{cccc}
      P_{qq} & 2n_{\F}P_{gq} & P_{sq} & 2n_{\F}P_{\lambda q} \\
      P_{qg} & P_{gg} & P_{sg} & P_{\lambda g}  \\
    P_{qs} & 2n_{\F}P_{gs} & P_{ss} & 2n_{\F}P_{\lambda s} \\
    P_{q\lambda} & P_{g\lambda} & P_{s\lambda} & P_{\lambda\lambda}
    \end{array}
  \right)
  \otimes
  \left(
    \begin{array}{l}
      D_q \\
      D_g \\
      D_s \\
      D_\lambda
    \end{array}
  \right). 
\end{equation}

We have developed a {\tt C++} code to solve the DGLAP equations for
fragmentation functions. We have extended the algorithm introduced in
Ref.~\cite{FurmanskiPetronzio} to be able to solve the SUSY equations
as well as the Standard Model equations~\cite{Toldra}.

We begin with FFs at the energy scale $\mu_0=M_Z$ and evolve them to
the final energy scale $\mu=M_X$ using Eq.~(\ref{eq:SM2X2}) and also
Eq~(\ref{eq:SUSY4X4}) if SUSY is included. For the initial
fragmentation function of baryons, $D^p_a(x,M_Z^2)+D^n_a(x,M_Z^2)$, we
adopt the fit performed in Ref.~\cite{Rubin} to LEP hadronic data
\cite{LEP}. For photons and neutrinos we generate initial data at the
$Z$ peak using the QCD Monte Carlo event generator HERWIG
\cite{HERWIG}. Comparison with LEP data shows that although HERWIG
overproduces baryons at high $x$ \cite{Rubin,kupco}, its photon and
meson output at the $Z$ peak matches the experimental spectra
remarkably well. Since neutrinos mainly come from charged pion and
kaon decays, one can thus be confident in taking the HERWIG generated
FF for neutrinos as the initial condition for the evolution. There is
also a sizable contribution from heavy flavour decays to the neutrino
spectrum at high $x$ which is explicitly taken into account by HERWIG.

A (s)parton is not included in the evolution as long as the energy
scale is lower than its mass; when the threshold for its production is
crossed, it is added to the evolution equations with an initially {\em
  vanishing} FF and it is assumed to be a relativistic particle.

In the SM case we evolve the $q$ and $g$ initial fragmentation
functions from $M_Z$ to $M_t$, the top quark mass, with the number of
flavours set to $n_\F=5$, and then evolve from $M_t$ to $M_X$ with
$n_\F=6$. 

In the SUSY case we evolve the $q$ and $g$ initial fragmentation
functions from $M_Z$ to the supersymmetry breaking scale $M_\SUSY>M_t$
using the SM equations to obtain $D^h_i(x,M^2_\SUSY)$, with $i=q,g$.
Then we take $D^h_i(x,M^2_\SUSY)$, $i=q,g$, and
$D^h_j(x,M^2_\SUSY)=0$, $j=s,\lambda$, and evolve them from $M_\SUSY$
to $M_X$ using the SUSY equations. All spartons are taken to be
degenerate with a common mass $M_\SUSY$.

\section{Ultra High Energy Cosmic Ray Spectrum}

We can now translate the calculated fragmentation functions into the
expected cosmic ray spectrum in order to confront the observational
data. In the previous Section (\ref{FFEvolution}) we have briefly
shown how to calculate quark singlet and gluon functions for the SM,
and quark singlet, gluon, squark singlet and gluino functions for
SUSY. In the absence of a specific model for the different branching
ratios we weight all the (s)parton contributions evenly.  For
2-body decay of $X$, $x=2E/M_X$ the flux of particle $h$ is given
by~\cite{SarkarToldra}:
\begin{equation}
  \label{eq:E3Flux}
  E^3 J^\halo (E) = B x^3 D^h (x, M^2_X).
\end{equation}
We have multiplied the flux by $E^3$, as is usual, to emphasise the
structure in the spectrum near the GZK energy. The normalisation
factor $B$ is common for the galactic halo flux of baryons, neutrinos
and photons, and determines the quantity $n_X/\tau_X$ (see
Eq.~\ref{eq:DecayRate}).

\begin{figure}[htb]
  \centering
  \includegraphics[height=4.3in]{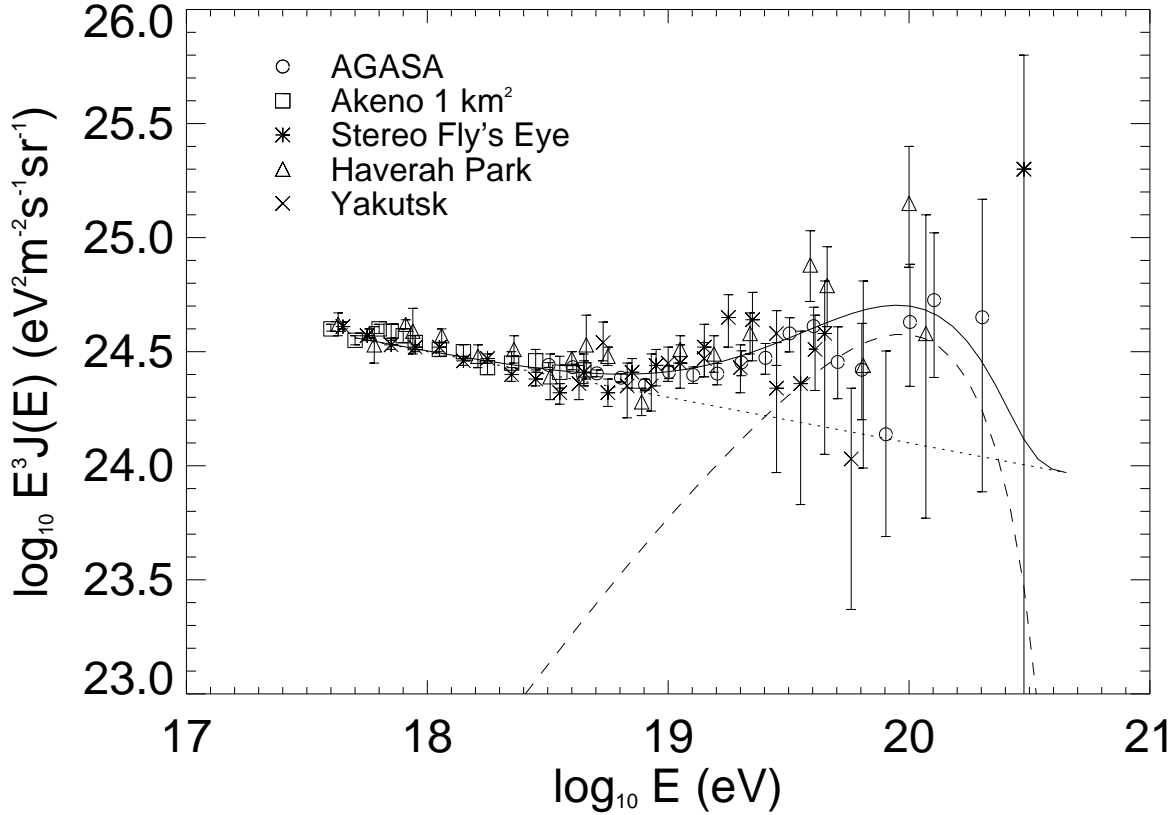}
  \caption{The best SM evolution fit to the cosmic ray data with a
  decaying particle mass of $10^{12}$~GeV. The dotted line indicates
  the extrapolation of the power-law component from lower energies,
  while the dashed line shows the decay spectrum; the solid line is
  their sum.}
  \label{fig:SMfit} 
\end{figure}

\begin{figure}[htb]
  \centering
  \includegraphics[height=4.3in]{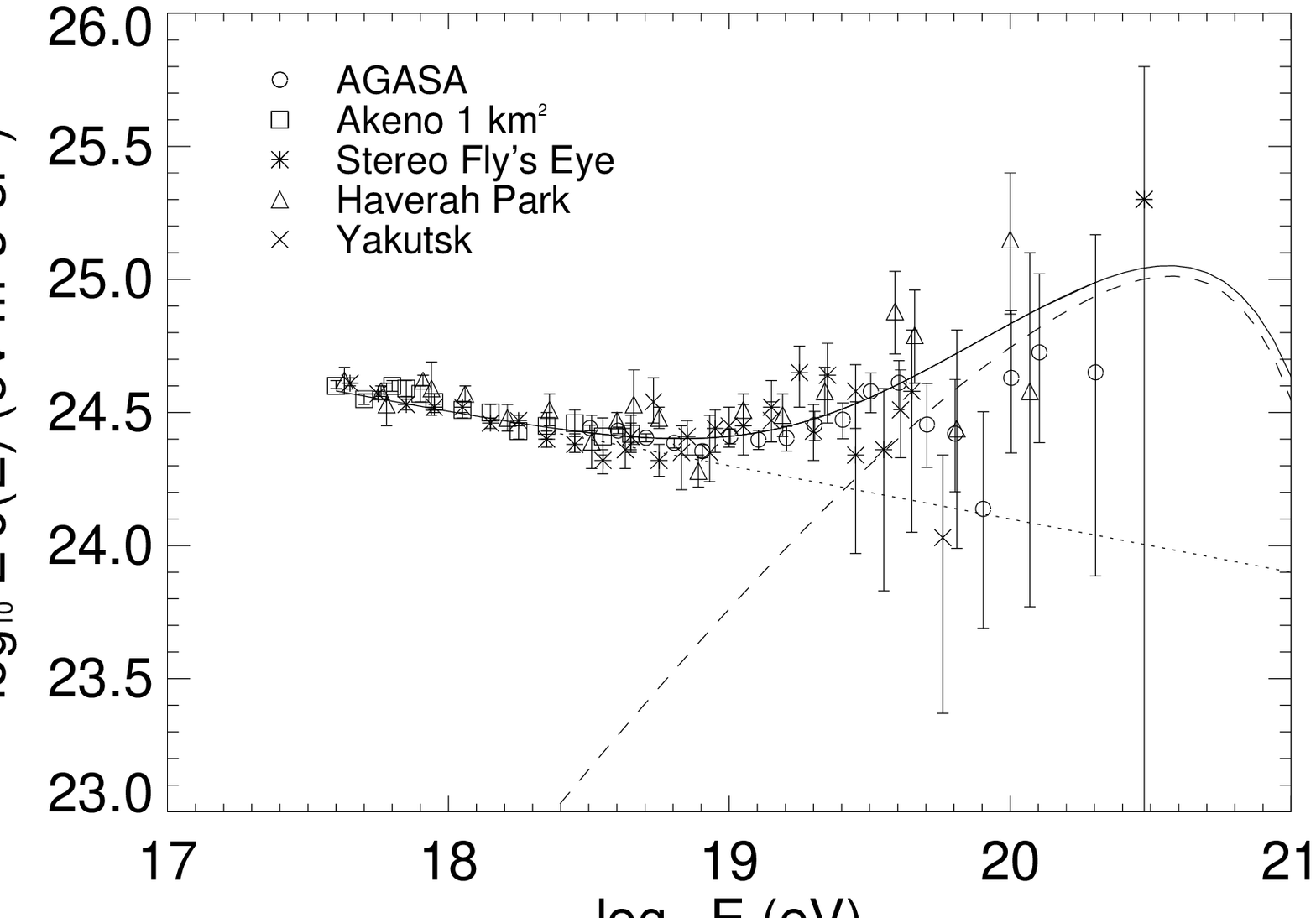}
  \caption{The best SUSY evolution fit to the cosmic ray data with a
    decaying particle mass of $5\times10^{12}$~GeV and
    $M_{SUSY}=400$~GeV. The dotted line indicates the extrapolation of
    the power-law component from lower energies, while the dashed line
    shows the decay spectrum; the solid line is their sum.}
  \label{fig:SUSYfit} 
\end{figure}

Let us now compare the calculated cosmic ray flux to the published
data from Fly's Eye \cite{fe}, AGASA \cite{agasa}, Haverah Park
\cite{hp} and Yakutsk \cite{yak}. In Ref.~\cite{NaganoWatson} these
data have been carefully assessed for mutual consistency and
appropriate adjustments made to the energy calibration. Its authors
recommend adoption of the following standard differential energy
spectrum {\em below} the GZK energy, in the range
$4\times10^{17}\ev<E<6.3\times10^{18}\ev$:
\begin{equation}
 \label{eq:LowComp}
  J(E) = (9.23\pm0.65)\times10^{-33}\;
  \mbox{m}^{-2}\mbox{s}^{-1}\mbox{sr}^{-1}\mbox{eV}^{-1}
  \left(\frac{E}{6.3\times10^{18}\;\mbox{eV}}\right)^{-3.20\pm0.05},
\end{equation}
with the spectrum flattening at higher energies as
$J(E)\propto\,E^{-2.75\pm0.2}$ up to the GZK energy, and extending
further to at least $3\times10^{20}\ev$ \cite{NaganoWatson}. Thus the
UHECR spectrum can naturally be interpreted \cite{fe} as the
superposition of the `low energy' component (\ref{eq:LowComp}), and
the new `flat' component that extends into the post-GZK region. The
former is presumably galactic in origin (consistent with the detection
of anisotropy at $\sim10^{18}\ev$ \cite{anisofe,diskaniso}), while the
latter is interpreted \cite{bs98} as produced by the decay of a
superheavy particle population in the galactic halo. Taking baryons to
be the dominant primary UHECRs as indicated by experiment, the total
flux is
\begin{equation}
  \label{eq:TotalFlux}
  E^3 J(E) = \frac{k}{E^m} + B x^3 D^\baryon (x,M^2_X),
\end{equation}
where the values of $k$ and $m$ can be read off
Eq.~(\ref{eq:LowComp}). Note that since $D^\baryon$ and $D^\gamma$
have a similar shape, taking photons to be the primaries would just
alter the normalisation $B$. In Fig.~\ref{fig:SMfit} we plot the best
SM evolution fit to the cosmic ray data while in
Fig.~\ref{fig:SUSYfit} we plot the best SUSY evolution fit.

The assumption of 2-body decay may be rather naive for a superheavy
particle like a crypton~\cite{crypton} which is expected to decay
through very high-order non-renormalisable operators. Many-body decay
distributes the total energy $M_X$ among several particles and thus
flattens the spectrum. We assume that many-body effects are purely
kinematical and hence can be encapsulated in the phase space of the
decay. Let $\rho_n(z)$ be the probability density that one parton
carries off a fraction $z$ of the total available energy per parton
$M_X/2$. For $n\ge3$ we get~\cite{SarkarToldra}
\begin{equation}
  \label{eq:rho3}
  \rho_n(z) \propto z (1-z)^{n-3}.
\end{equation}
To leading order in $\alpha_s$ the particle flux is then given by
\begin{equation}
 \label{eq:E3FluxMultiBody}
  E^3 J^\halo (E)= B x^3 \int^1_x \frac{\rmd z}{z} 
  \rho_n\left(\frac{x}{z}\right)  D^h (z, M^2_X).
\end{equation}
In particular if $D^h(x,M^2_X)\propto\,(1-x)^{a(M^2_X)}$ as
$x\rightarrow1$, the differential particle flux decreases as
$J^\halo(x)\propto(1-x)^{a(M^2_X)+n-2}$.

\begin{figure}[htb]
  \centering
  \includegraphics[height=4.3in]{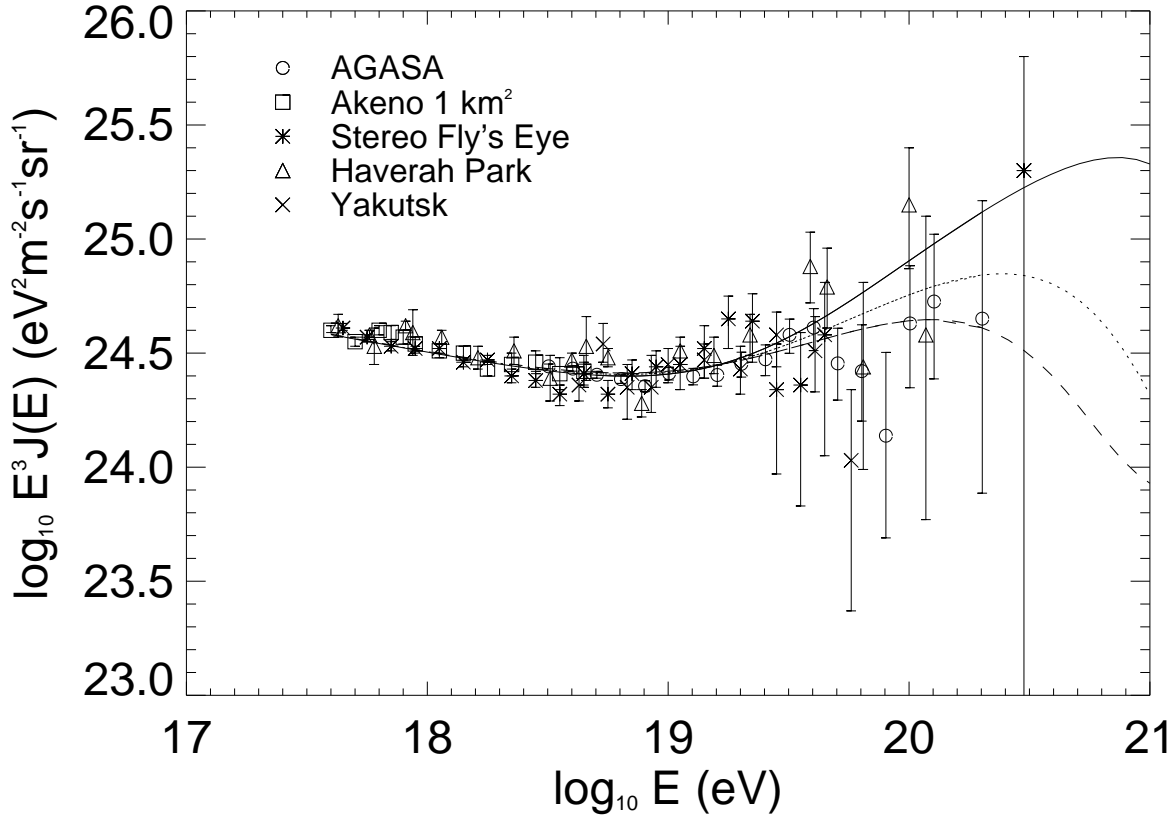}    
  \caption{Cosmic ray data compared with SUSY evolved spectra for a
  decaying particle mass of $10^{13}$~GeV with $M_\SUSY=400$~GeV,
  and many-body decays to $n$ partons: $n=2$ (solid line), $n=8$
  (dotted line) and $n=16$ (dashed line).}  
  \label{fig:nbody}
\end{figure}

For many-body $X$ decays the total flux is
\begin{equation}
  \label{eq:TotalFluxMultiBody}
   E^3 J(E) = \frac{k}{E^m} +
   B x^3 \int^1_x \frac{\rmd z}{z} \rho_n\left(\frac{x}{z}\right) 
   D^\baryon (z , M^2_X),
\end{equation}
where $n$ is the number of partons into which $X$ decays. In
Fig.~\ref{fig:nbody} we plot evolved SUSY spectra for different
values~of~$n$.

\section{Conclusions}

We have calculated the UHECR spectra of baryons, photons and neutrinos
expected from the decay of superheavy dark matter particles of mass
$M_X\sim 10^{12}$~GeV which are clustered in the galactic halo. The
spectrum for every primary particle is given by FFs at the scale
$M_X$. We have calculated these FFs using QCD DGLAP
equations. We have considered Standard Model equations and SUSY
equations. We have also taken into account the possibility of
many-body decay in the decay of $X$. The shape of the fragmentation
spectrum (of either baryons or photons) fits rather well the new
component of ultra-high energy cosmic rays extending beyond the GZK
energy.












\Acknowledgements

The work here outlined was performed in collaboration with Subir
Sarkar. I would like to acknowledge support by a Marie Curie
Fellowship No.~HPMF-CT-1999-00268.

\end{document}